\documentclass[11pt]{article}

\usepackage{ifthen}
\usepackage{mdwlist}
\usepackage{fullpage}
\usepackage{amsmath,amsfonts,amsthm}
\usepackage{enumitem}
\usepackage{graphicx}
\usepackage[ruled,linesnumbered]{algorithm2e}

\def\eps{\epsilon}

\def\to{\rightarrow}
\def\E{{\bf E}}

\def\calt{{\cal T}}

\def\msp{{\Obj}}
\def\Emsp{{\overline{\Obj}}}
\def\Feas{{\cal F}}
\def\Obj{{\phi}}

\newcommand{\comment}[1]{}

\newcommand{\prob}[2][]{\text{\bf Pr}\ifthenelse{\not\equal{}{#1}}{_{#1}}{}\!\left[#2\right]}
\newcommand{\expect}[2][]{\text{\bf E}\ifthenelse{\not\equal{}{#1}}{_{#1}}{}\!\left[#2\right]}

\newcommand{\reals}{\mathbb{R}}

\newtheorem{thm}{Theorem}[section]

\newtheorem{lem}[thm]{Lemma}

\newtheorem{claim}[thm]{Claim}

\newtheorem{obs}[thm]{Observation}

\DeclareMathOperator{\argmax}{argmax}

\newcommand{\bg}[1]{\medskip\noindent{\bf #1}}
\newenvironment{proofof}[1]{\bg{Proof of #1 : }}{\qed}

\newcommand{\alg}{{\cal A}}

\newcommand{\bid}{b}
\newcommand{\bids}{{\mathbf \bid}}
\newcommand{\bidsmi}{{\mathbf \bid}_{-i}}
\newcommand{\bidi}[1][i]{{\bid_{#1}}}

\newcommand{\val}{v}
\newcommand{\vals}{{\mathbf \val}}
\newcommand{\valsmi}{\vals_{-i}}
\newcommand{\vali}[1][i]{{\val_{#1}}}

\newcommand{\dist}{F}
\newcommand{\dists}{{\mathbf \dist}}

\newcommand{\disti}[1][i]{{\dist_{#1}}}

\newcommand{\price}{p}
\newcommand{\prices}{{\mathbf \price}}
\newcommand{\pricei}[1][i]{{\price_{#1}}}

\newcommand{\alloc}{x}
\newcommand{\allocs}{{\mathbf \alloc}}

\newcommand{\alloci}[1][i]{{\alloc_{#1}}}

\newcommand{\approxratio}{\beta}

\title{On the Impossibility of Black-Box Transformations in Mechanism Design}
\author{
Shuchi Chawla\thanks{Department of Computer Science, University of Wisconsin - Madison.} 
\and
Nicole Immorlica\thanks{Department of Electrical Engineering and Computer Science, Northwestern University.} 
\and 
Brendan Lucier\thanks{Department of Computer Science, University of Toronto.}
}
\date{}

\begin{document}
\setcounter{page}{0}


\maketitle

\begin{abstract}
  We consider the problem of converting an arbitrary approximation
  algorithm for a single-parameter optimization problem into a
  computationally efficient truthful mechanism. We ask for reductions
  that are black-box, meaning that they require only oracle access to
  the given algorithm and in particular do not require explicit
  knowledge of the problem constraints. Such a reduction is known to
  be possible, for example, for the social welfare objective when the
  goal is to achieve Bayesian truthfulness and preserve social welfare
  in expectation. We show that a black-box reduction for the social
  welfare objective is not possible if the resulting mechanism is
  required to be truthful in expectation and to preserve the
  worst-case approximation ratio of the algorithm to within a
  subpolynomial factor. Further, we prove that for other objectives
  such as makespan, no black-box reduction is possible even if we only
  require Bayesian truthfulness and an average-case performance
  guarantee.
\end{abstract}



\thispagestyle{empty}

\newpage

\section{Introduction}

Mechanism design studies optimization problems arising in settings
involving selfish agents, with the goal of designing a system or
protocol whereby agents' individual selfish optimization leads to
global optimization of a desired objective. A central theme in {\em
  algorithmic} mechanism design is to reconcile the incentive
constraints of selfish participants with the requirement of
computational tractability and to understand whether the combination
of these two considerations limits algorithm design in a way that each
one alone does not.

In the best-case scenario, one might hope for a sort of equivalence
between the considerations of algorithm design and mechanism design.
In particular, recent research has explored general reductions that
convert an arbitrary approximation algorithm into an incentive
compatible mechanism.  Ideally, these reductions have an arbitrarily
small loss in performance, and are black-box in the sense that they
need not understand the underlying structure of the given algorithm or
problem constraints. A big benefit of this approach is that it allows
a practitioner to ignore incentive constraints while fine-tuning his
algorithm to the observed workload. Of course the feasibility of the
approach depends heavily on the objective at hand as well as the
incentive requirements. The goal of this paper is to understand what
scenarios enable such black-box reductions. 

The classic result of Vickrey, Clarke and Groves provides a positive result along these lines for social welfare maximization. Social welfare maximization is a standard objective in
mechanism design. Here, a central authority wishes to assist a group
of individuals in choosing from among a set of outcomes with the goal
of maximizing the total outcome value to all participants.  The
Vickrey, Clarke and Groves result demonstrates that for any
such problem, there exists a mechanism that maximizes social welfare
with very robust incentive properties (namely, it is {\it ex post
  incentive compatible}).  This construction requires that the
mechanism optimize social welfare precisely, and so can be thought of
as a reduction from incentive compatible mechanism design to exact
algorithm design.  
This result can be extended to more general scenarios beyond social welfare.  In the single-parameter setting, where the preferences of every selfish agent can
be described by a single scalar parameter, (Bayesian or ex post)
incentive compatibility is essentially 
equivalent to a per-agent
monotonicity condition on the allocation returned by the mechanism. 
Therefore, for objective functions that are ``monotone'' in the sense that exact
optimization of the objective leads to a monotone allocation
function, there is a reduction from mechanism design to
exact algorithm design along the lines of the VCG mechanism for social
welfare.

However, many settings of interest involve constraints
which are computationally infeasible to optimize precisely, and so exact algorithms are not known to exist.
Recent
work~\cite{HL-10, HKM-11, BH-11} shows that in Bayesian settings of
partial information, the reduction from mechanism design to algorithm
design for social welfare can be extended to encompass arbitrary
approximate algorithms with arbitrarily small loss in expected
performance. These reductions work even in multi-parameter settings.  Moreover, these reductions are black-box, meaning that
they require only oracle access to the prior type distributions and
the algorithm, and proceed without knowledge of the feasibility
constraints of the problem to be solved.

In light of this positive result, two natural questions arise:
\begin{itemize*}
\item Are there 																	 black-box reductions transforming approximation
  algorithms for social welfare into {\em ex post} incentive compatible
  mechanisms with little loss in {\em worst-case} approximation ratio?
\item Does {\em every} monotone objective 
  admit a black-box
  reduction that transforms approximation algorithms into (Bayesian or
  ex post) incentive compatible mechanisms with little loss in (worst-case or average) approximation ratio?
\end{itemize*}

\noindent
In this paper we answer both of these questions in the negative. Our
impossibility results apply to the simpler class of single-parameter
optimization problems.

The first question strengthens the demands of the reduction beyond those of the aforementioned positive results~\cite{HL-10, HKM-11, BH-11} in two
significant ways. First, it requires the stronger solution concept of
ex post incentive compatibility, rather than Bayesian incentive
compatibility.  Second, it requires that the approximation factor of
the original algorithm be preserved in the worst case over all
possible inputs, rather than in expectation. It is already known that
such reductions are not possible for general multi-parameter social
welfare problems.  For some multi-parameter problems, no
ex post incentive compatible mechanism can match the worst-case
approximation factors achievable by algorithms without game-theoretic
constraints~\cite{PSS-08, DOB-11}.  Thus, for general social welfare problems,
the relaxation from an ex post, worst-case setting to a Bayesian
setting provably improves one's ability to implement algorithms as
mechanisms. However, prior to our work, it was not known whether a
lossless black-box reduction could exist for the important special case of 
single-parameter problems. We show that any such reduction for
single-parameter problems must sometimes degrade an algorithm's 
worst-case performance by a factor that is polynomial in the problem size.

The second question asks whether there are properties specific to
social welfare that enable the computationally efficient reductions of
the above results~\cite{HL-10, HKM-11, BH-11}. One property that appears to be crucial in
their analysis is the linearity of the objective in both the agents'
values and the algorithm's allocations. For our second impossibility
result we consider the (highly non-linear but monotone) makespan
objective for scheduling problems. In a scheduling problem we are
given a number of (selfish) machines and jobs; our goal is to schedule
the jobs on the machines in such a way that the load on the most
loaded machine (namely, the makespan of the schedule) is
minimized. The sizes of jobs on machines are private information that
the machines possess and must be incentivized to share with the
algorithm.

Ashlagi et al.~\cite{ADL09} showed that for the makespan objective in
multi-parameter settings (that is, when the sizes of jobs on different
machines are unrelated), ex post incentive compatibility imposes a
huge cost: while constant
factor approximations can be obtained in the absence of incentive constraints, 
no ``anonymous'' mechanism can obtain a
sublinear approximation ratio under the requirement of
incentive compatibility. The situation for single-parameter
settings is quite different. In single-parameter (a.k.a. related)
settings, each machine has a single private parameter, namely its
speed, and each job has a known intrinsic size; the load that a job
places on a machine is its size divided by the speed of the
machine. In such settings in the absence of feasibility constraints,
deterministic PTASes are known both with and without incentives. Given
this positive result one might expect that for single-parameter
makespan minimization there is no gap between algorithm design and
mechanism design, at least in the weaker Bayesian setting where the
goal is to achieve Bayesian incentive compatibility and match the
algorithm's expected performance. We show that this is not true---any
black-box reduction that achieves Bayesian incentive compatibility
must sometimes make the expected makespan worse by a factor polynomial in the
problem size.

Finally, while makespan is quite different from social welfare, one
might ask whether there exist objectives that share some of the nice
properties of social welfare that enable reductions in the style of
\cite{HKM-11} and others. At a high level, the black-box reductions
for social welfare perform``ironing'' operations for each agent
independently fixing non-monotonicities in the algorithm's output in a
local fashion without hurting the overall social welfare. One property
of social welfare that enables such an approach is that it is additive
across agents. In our final result we show that even restricting
attention to objectives that are additive across agents, for almost
any objective other than social welfare no per-agent ironing procedure
can simultaneously ensure Bayesian incentive compatibility as well as
a bounded loss in performance. The implication for mechanism design is
that any successful reduction must take a holistic approach over
agents and look very different from those known for social welfare.

\paragraph{Our results and techniques.}
As mentioned earlier, the existence of a black-box reduction from
mechanism design to algorithm design can depend on the objective
function we are optimizing, the incentive requirements, as well as
whether we are interested in a worst-case or average-case performance
guarantee. We distinguish between two kinds of incentive requirements
(see formal definitions in Section~\ref{sec.preliminaries}). Bayesian incentive
compatibility (BIC) implies that truthtelling forms a Bayes-Nash
equilibrium under the assumption that the agents' value distributions
are common knowledge. The stronger notion of ex post incentive
compatibility (EPIC), a.k.a. universal truthfulness, implies that
every agent maximizes her utility by truthtelling regardless of
others' actions and the mechanism's coin flips. For randomized
mechanisms there is a weaker notion of truthfulness called
truthfulness in expectation (TIE) which implies that every agent
maximizes her utility in expectation over the randomness in the
mechanism by truthtelling regardless of others' actions. We further
distinguish between social welfare and arbitrary monotone objectives,
and between the average performance of the algorithm and its worst
case performance.

Table~\ref{tab:results} below summarizes our findings as well as known
results along these three dimensions. Essentially, we find that there
is a dichotomy of settings: some allow for essentially lossless
transformations whereas others suffer an unbounded loss in
performance.

\begin{table}[h]
\begin{small}
\begin{center}
\begin{tabular}{cc}

Objective: social welfare
&
Objective: arbitrary monotone (e.g. makespan)\\

\begin{tabular}{|l|c|c|}
\hline
& Avg-case approx & Worst-case approx \\
\hline
\hline
BIC & Yes \cite{HL-10, BH-11, HKM-11}& ?\\
\hline
TIE & ? & No (Section~\ref{sec.welfare}) \\
\hline
\end{tabular}

&

\begin{tabular}{|l|c|c|}
\hline
& Avg-case approx & Worst-case approx \\
\hline
\hline
BIC & No (Section~\ref{sec.makespan}) & No \\
\hline
TIE & No & No \\
\hline
\end{tabular}

\end{tabular}
\end{center}
\caption{\label{tab:results} A summary of our results on the existence
of black-box transformations. A ``yes'' indicates that a reduction
exists and gives an arbitrarily small loss in performance; a ``no''
indicates that every reduction satisfying incentive constraints
suffers an arbitrarily large loss in performance.}
\end{small}
\end{table}

One way to establish our impossibility results would be to demonstrate
the existence of single-parameter optimization problems for which
there is a gap in the approximating power of arbitrary algorithms and
ex post incentive compatible algorithms.  This is an important open
problem which has resisted much effort by the algorithmic mechanism
design community, and is beyond the scope of our work.
Instead, we focus upon the black-box nature of the reductions with
respect to, in particular, the feasibility constraint that they
face. Note that for single-parameter problems, (Bayesian or ex post)
incentive compatibility is essentially equivalent to a per-agent
monotonicity condition on the allocation returned by the mechanism. We
construct instances that contain ``hidden'' non-monotonicities and yet
provide good approximations.  In order for the transformation to be
incentive compatible while also preserving the algorithm's
approximation factor, it must fix these non-monotonicities by
replacing the algorithm's original allocation with very specific kinds
of ``good'' allocations. However, in order to determine which of these good
allocations are also feasible the transformation must query the
original algorithm at multiple inputs with the hope of finding a good
allocation. We construct the algorithm in such a way that any single
query of the transformation is exponentially unlikely to find a good
allocation. 

\paragraph{Related Work.}

Reductions from mechanism design to algorithm design in the Bayesian setting
were first studied by Hartline and Lucier \cite{HL-10}, who showed that
any approximation algorithm for a single-parameter social welfare problem
can be converted into a Bayesian incentive
compatible mechanism with arbitrarily small loss in expected performance.
This was extended to multi-parameter settings by Hartline, Kleinberg and 
Malekian \cite{HKM-11} and Bei and Huang \cite{BH-11}.

Some reductions from mechanism design to algorithm design are 
known for prior-free settings, for certain restricted classes of algorithms.  
Lavi and Swamy \cite{LS-05} consider mechanisms for multi-parameter
packing problems and show how to construct a (randomized) 
$\approxratio$-approximation mechanism that is truthful in
expectation, from any $\approxratio$-approximation that 
verifies an integrality gap.  Dughmi, Roughgarden and Yan \cite{DRY-11}
extend the notion of designing mechanisms based upon randomized rounding
algorithms, and obtain truthful in expectation mechanisms for a broad class
of submodular combinatorial auctions.
Dughmi and Roughgarden \cite{DR-10} give a construction that converts
any FPTAS algorithm for a social welfare problem into a mechanism that is
truthful in expectation, by way of a variation on smoothed analysis.

Babaioff et al.~\cite{BLP-09} provide a
technique for turning a $\approxratio$-algorithm for a single-valued
combinatorial auction problem into a
truthful $\approxratio (\log \val_{max}/\val_{min})$-approximation mechanism,
when agent values are restricted to lie in $[\val_{min},\val_{max}]$.
This reduction applies to single-parameter problems with downward-closed
feasibility constraints and binary allocations (each agent's allocation
can be either $0$ or $1$).

Many recent papers have explored limitations on the power of 
deterministic ex post incentive compatible mechanisms to approximate 
social welfare.  Papadimitriou, Schapira and Singer \cite{PSS-08} gave an
example of a social welfare problem for which constant-factor approximation
algorithms exist, but any polytime ex post incentive compatible mechanism 
attains at best a polynomial approximation factor.  A similar gap for the 
submodular combinatorial auction problem was established by Dobzinski 
\cite{DOB-11}.  For the general combinatorial auction problem, such gaps 
have been established for the restricted
class of max-in-range mechanisms by Buchfuhrer et al. \cite{BDFKMPSSU-10}.

Truthful scheduling on related machines to minimize makespan was studied by
Archer and Tardos \cite{AT-01}, who designed a truthful-in-expectation 3-approximation.
Dhangwatnotai et al. \cite{DDDR-08} gave a randomized PTAS that is truthful-in-expectation, 
which was then improved to a deterministic truthful PTAS by Christodoulou and 
Kov\'{a}cs \cite{CK-10}, matching
the performance of the best possible approximation algorithm \cite{HS-88}. 
Our work on makespan minimization differs in that we 
consider the goal of minimizing makespan subject to an arbitrary feasibility constraint.

A preliminary version of this work \cite{IL-11} proved an impossibility 
result for EPIC black-box reductions for single-parameter social welfare 
problems.  In this paper we extend that result to apply to (the broader 
class of) TIE reductions.

\section{Preliminaries}
\label{sec.preliminaries}

\paragraph{Optimization Problems.}

In a single-parameter real-valued optimization
problem we are given an input vector $\vals = (\vali[1], \vali[2],
\dotsc, \vali[n])$.  Each $\vali$ is assumed to be drawn from a
known set $V_i \subseteq \reals$, so that $V = V_1 \times \cdots \times V_n$
is the set of possible input vectors.  The goal is to choose some allocation $\allocs
\in \Feas \subseteq \reals^n$ from among a set of feasible allocations $\Feas$ such that a given objective function 
$\Obj : \Feas \times V \to \reals$ is optimized (i.e.\ either maximized or 
minimized, depending on the nature of the 
problem).  We think of the feasibility set $\Feas$ and the objective function
$\Obj$ as defining an instance of the optimization problem.  We will write
$\allocs = (\alloci[1], \alloci[2], \dotsc, \alloci[n])$, where each
$\alloci \in \reals$.


An algorithm $\alg$ defines a mapping from input vectors $\vals$ to
outcomes $\allocs$.  We will write $\alg(\vals)$ for the allocation
returned by $\alg$ as well as the value it obtains; the intended
meaning should be clear from the context.  In general an algorithm can
be randomized, in which case $\alg(\vals)$ is a random variable.

Given an instance $\Feas$ of the social welfare problem, we will write
$OPT_{\Feas}(\vals)$ for the allocation in $\Feas$ that maximizes
$\Obj(\allocs,\vals)$, as well as the value it obtains.  Given
algorithm $\alg$, let $approx_{\Feas}(\alg)$ denote the worst-case
approximation ratio of $\alg$ for problem $\Feas$.  That is,
$approx_{\Feas}(\alg) = \min_{\vals \in V}
\frac{\alg(\vals)}{OPT_\Feas(\vals)}$ for a maximization problem; here
$\Obj$ is implicit and should be clear from the context.  Note that
$approx_{\Feas}(\alg) \leq 1$ for all $\Feas$ and $\alg$.

We also consider a Bayesian version of our optimization problem,
in which there is publicly-known product distribution $\dists$ on 
input vectors.  That is, $\dists = \disti[1] \times \dotsc \times \disti[n]$
and each $\vali$ is distributed according to $\disti$.
Given $\dists$, the expected
objective value of a given algorithm $\alg$ is given by
$\overline{\Obj}(\alg)=\E_{\vals \sim \dists}[\Obj(\alg(\vals), \vals)]$.  The goal of the
optimization problem in this setting is to optimize the expected objective 
value.  


\paragraph{Mechanisms.}

We will consider our optimization problems in a
mechanism design setting with $n$ rational agents, where each agent
possesses one value from the input vector as private information.  We
think of an outcome $\allocs$ as representing an \emph{allocation} to
the agents, where $\alloci$ is the allocation to agent $i$.  A
(direct-revelation) mechanism for our optimization problem then
proceeds by eliciting declared values $\bids \in \reals^n$ from the
agents, then applying an allocation algorithm $\alg : \reals^n \to \Feas$
that maps $\bids$ to an allocation $\allocs$, and a payment rule that
maps $\bids$ to a payment vector $\prices$.  We will write
$\allocs(\bids)$ and $\prices(\bids)$ for the allocations and payments
that result on input $\bids$.  The \emph{utility} of agent $i$, given
that the agents declare $\bids$ and his true private value is $\vali$,
is taken to be $\vali \alloci(\bids) - \pricei(\bids)$.

A (possibly randomized) mechanism is \emph{truthful in expectation} (TIE) if each
agent maximizes its expected utility by reporting its value truthfully,
regardless of the reports of the other agents, where expectation
is taken over any randomness in the mechanism.  That is, $\E[\vali
\alloci(\vali, \bidsmi) - \pricei(\vali, \bidsmi)] \geq \E[\vali
\alloci(\bidi, \bidsmi) - \pricei(\bidi, \bidsmi)]$ for all $i$, all
$\vali, \bidi \in V_i$, and all $\bidsmi \in V_{-i}$.  We say that
an algorithm is TIE if there exists a
payment rule such that the resulting mechanism is TIE.  It is
known that an algorithm is TIE if and only if, for all $i$ and 
all $\valsmi$, $\E[\alloci(\vali, \valsmi)]$ is monotone non-decreasing 
as a function of $\vali$, where the expectation is over the randomness in the mechanism.

We say that a (possibly randomized) mechanism is
\emph{Bayesian incentive compatible} (BIC) for distribution $\dists$
if each agent maximizes its expected utility by reporting its value truthfully,
given that the other agents' values are distributed according to $\dists$
(and given any randomness in the mechanism).
That is, $\E_{\valsmi}[\vali
\alloci(\vali, \valsmi) - \pricei(\vali, \valsmi)] \geq \E_{\valsmi}[\vali
\alloci(\bidi, \valsmi) - \pricei(\bidi, \valsmi)]$ for all $i$ and all
$\vali, \bidi \in V_i$, where the expectation is over the distribution of others' values and the randomness in the mechanism.  We say that an algorithm is BIC if there exists a
payment rule such that the resulting mechanism is BIC.  It is
known that an algorithm is BIC if and only if, for all $i$, 
$\E_{\valsmi}[\alloci(\vali, \valsmi)]$ is monotone non-decreasing 
as a function of $\vali$.

\paragraph{Transformations.}

A \emph{polytime transformation} $\calt$ is an algorithm
that is given black-box access to an algorithm $\alg$.  We will write
$\calt(\alg,\vals)$ for the allocation returned by $\calt$ on input
$\vals$, given that its black-box access is to algorithm $\alg$.
Then, for any $\alg$, we can think of $\calt(\alg, \cdot)$ as an
algorithm that maps value vectors to allocations; we think of this as the
algorithm $\alg$ \emph{transformed by $\calt$}.  We write
$\calt(\alg)$ for the allocation rule that results when $\alg$
is transformed by $\calt$.  Note that $\calt$ is not parameterized by
$\Feas$; informally speaking, $\calt$ has no knowledge of the
feasibility constraint $\Feas$ being optimized by a given algorithm
$\alg$.  However, we do assume that $\calt$ is aware of the 
objective function $\Obj$, the domain $V_i$ of values for each agent $i$,
and (in Bayesian settings) the distribution $\dists$ over values.

We say that a transformation $\calt$ is truthful in expectation (TIE) if,
for all $\alg$, $\calt(\alg)$ is a TIE algorithm.  
In a Bayesian setting with distribution $\dists$, we say that 
transformation $\calt$ is Bayesian incentive compatible (BIC) for
$\dists$ if, for all $\alg$, $\calt(\alg)$ is a BIC algorithm.  
Note that whether or
not $\calt$ is TIE or BIC is independent of the objective function $\Obj$ and
feasibility constraint $\Feas$.

\section{A Lower Bound for TIE Transformations for social welfare}
\label{sec.welfare}

In this section we consider the problem of maximizing social welfare.  For this problem, BIC transformations that approximately preserve expected performance are known to exist. We prove that if we strengthen our solution concept to truthfulness in expectation and our performance metric to worst-case approximation, then such black-box transformations are not possible.  

\subsection{Problem definition and main theorem}

The social welfare objective is defined as $\Obj(\allocs,\vals) = \vals \cdot \allocs$.

Our main result is that, for any TIE transformation $\calt$, there is a problem instance $\Feas$ and algorithm $\alg$ such that $\calt$ degrades the worst-case performance of $\alg$ by a polynomially large factor.

\begin{thm}
\label{thm.tie.lowerbound}
There is a constant $c > 0$ such that, for any polytime TIE transformation $\calt$, there is an algorithm $\alg$ and problem instance $\Feas$ such that $\frac{approx_\Feas(\alg)}{approx_\Feas(\calt(\alg))} \geq n^c$.
\end{thm}

The high-level idea behind our proof of Theorem
\ref{thm.tie.lowerbound} is as follows.  We will construct an
algorithm $\alg$ and input vectors $\vals$ and $\vals'$ such that, for
each agent $i$ in some large subset of the players, $\vali' > \vali$
but $\alg_i(\vals') < \alg_i(\vals)$.  This does not immediately imply
that $\alg$ is non-truthful, but we will show that it does imply
non-truthfulness under a certain feasibility condition $\Feas$, namely
that any allocation is constant on the players $i$ with
$\vali'>\vali$.  Thus, any TIE transformation $\calt$ must alter the
allocation of $\alg$ either on input $\vals$ or on input $\vals'$.
However, we will craft our algorithm in such a way that, on input
$\vals$, the only allocations that the transformation will observe
given polynomially many queries of $\alg$ will be $\alg(\vals)$, plus
allocations that have significantly worse social welfare than
$\alg(\vals)$, with high probability.  Similarly, on input $\vals'$,
with high probability the transformation will only observe allocation
$\alg(\vals')$ plus allocations that have significantly worse social
welfare than $\alg(\vals')$.  Furthermore, we ensure that the
transformation can not even find the magnitude of the allocation to
players $i$ in $\val'$ when presented with input $\val$, thereby
preventing the transformation from randomizing between the high
allocation of $\alg(\vals)$ and an essentially empty allocation to
simulate the $\alg(\vals')$ allocation without directly observing it.
Instead, in order to guarantee that it generates an TIE allocation
rule, the transformation will be forced to assume the worst-case and
offer players $i$ the smallest possible allocation on input $\vals$.
This signifcantly worsens the worst-case performance of the algorithm
$\alg$.

\subsection{Construction}

In the instances we consider, each private value $\vali$ is chosen from $\{v,1\}$, where $0 < v < 1$ is a parameter that we set below.  That is, we will set $V_i = \{v,1\}$ for all $i \in [n]$.  We can therefore interpret an input vector as a subset $y \subseteq [n]$, corresponding to those agents with value $1$ (the remaining agents have value $v$). Accordingly we define $\alg(y)$, $OPT_{\Feas}(y)$, etc., for a given subset $y \subseteq [n]$.  Also, for $a \geq 0$ and $y \subseteq [n]$, we will write $\allocs^a_y$ for the allocation in which each agent $i \in y$ is allocated $a$, and each agent $i \not\in y$ is allocated $0$.

%
%

\paragraph{Feasible Allocations.}
We now define a family of feasibility constraints.  Roughly
speaking, we will choose $\alpha,\gamma \in (0,1)$ with $\gamma < \alpha$ and sets $S, T
\subseteq [n]$ of agents.
The feasible allocations will be $\allocs_{[n]}^{\gamma}$, $\allocs_{S}^{1}$, and $\allocs_{T}^{\alpha}$.
That is, we can allocate $\gamma$ to every agent, $1$ to all agents in $S$, or $\alpha$ to all agents in $T$.
We will also require that $S$ and $T$ satisfy
certain properties, which essentially state that $S$ and $T$ are
sufficiently large and have a sufficiently large intersection.

More formally, define parameters $\gamma \in (0,1)$, $\alpha \in (0,1)$, $r \geq 1$, and $t \geq 1$ (which we will fix later to be functions of $n$), such that 
$t \gg r \gg \gamma^{-1} \gg \alpha^{-1}$, $r^5t \leq n$, and $\frac{t}{\gamma n} \ll 1$.
We think of $t$ as a bound on the size of ``small'' sets, and we think of $r$ as a ratio between the sizes of ``small'' and ``large'' sets.  

\begin{figure}
\begin{center}
\begin{tabular}{c|c}
\includegraphics[viewport=0.0in 4.5in 3.5in 7.5in,clip,scale=0.6]{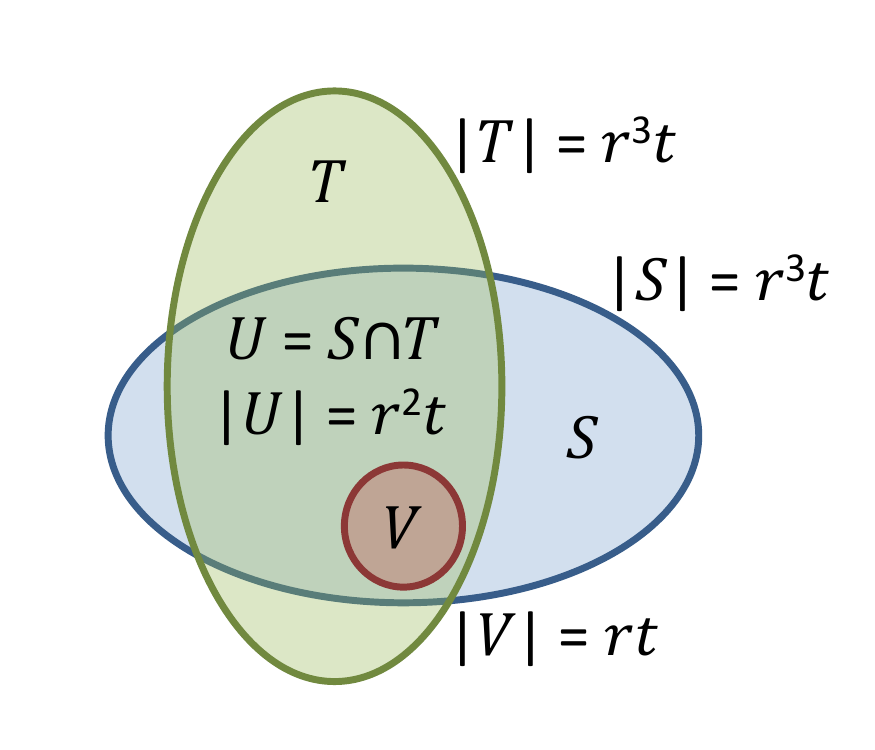} &
\includegraphics[viewport=0.0in 4.5in 3.5in 7.5in,clip,scale=0.6]{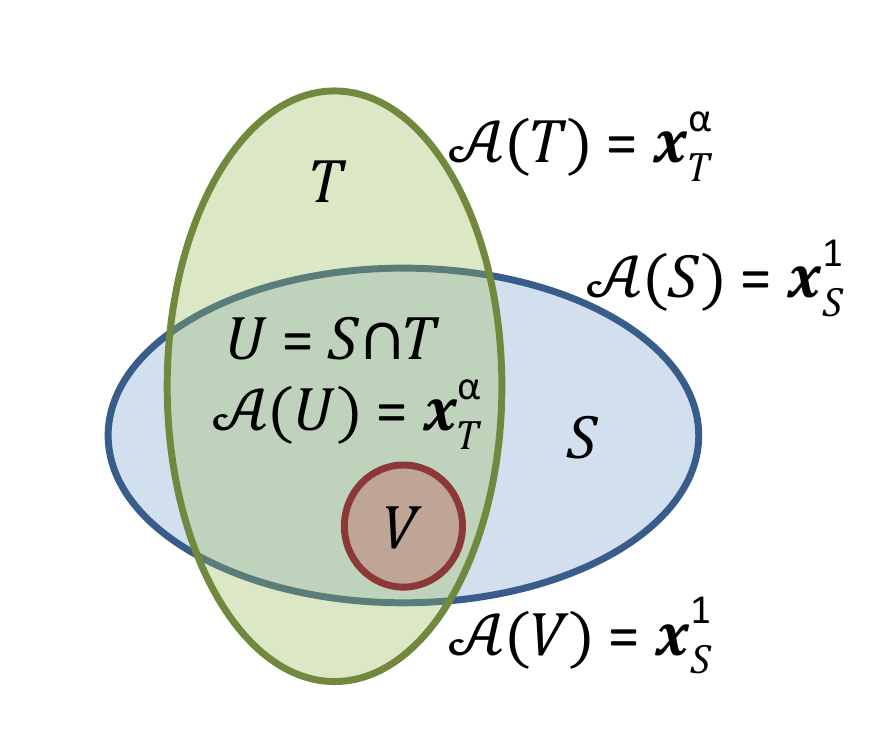} \\
(a) & (b) \\
\end{tabular}
\end{center}
\caption{\small (a) Visualization of typical admissible sets of bidders $V$, $S$,
  and $T$, given size parameters $r$ and $t$, and (b) the corresponding allocations of algorithm $\alg = \alg_{V,S,T,\alpha}$. }
\label{fig.lb1}
\end{figure}

Suppose that $V$, $S$, and $T$ are subsets of $[n]$.  We say that the
triple $V$, $S$, $T$ is \emph{admissible} if the following conditions
hold:
\begin{enumerate}
\item $|S| = |T| = r^3t$,
\item $|S \cap T| = r^2t$,
\item $V \subset S \cap T$, and,
\item $|V| = rt$.
\end{enumerate}
In general, for a given admissible $V$, $S$, and $T$, we will tend to
write $U = S \cap T$ for notational convenience.  See Figure
\ref{fig.lb1}(a) for an illustration of the relationship between the
sets in an admissible triple. In order to hide the feasibility
constraint $\Feas$ from the transformation, we will pick the sets $V$,
$S$, and $T$ uniformly at random from all admissible triples, and the
value $\alpha$ from an appropriate range.  For each admissible tuple
$V$, $S$, $T$, and value $\alpha$, we define a corresponding feasibility
constraint $$\Feas_{V,S,T,\alpha} = \{ x_S^1, x_T^\alpha,
x_{[n]}^{\gamma} \}.$$ Note that $\Feas_{V,S,T,\alpha}$ does not
depend on $V$; we include set $V$ purely for notational
convenience. We remark that all of the feasible allocations allocate
the same amount to agents in $U$.

Recall that agents have values chosen from $\{v,1\}$.  We will choose $v = \frac{t}{\gamma n}$, where we recall that our parameters have been chosen so that $\frac{t}{\gamma n} \ll 1$.

\paragraph{The Algorithm.}
We now define the algorithm $\alg_{V,S,T,\alpha}$ corresponding to an
admissible tuple $V, S, T$ and value $\alpha$.  We think of
$\alg_{V,S,T,\alpha}$ as an approximation algorithm for the social
welfare problem $\Feas_{V,S,T,\alpha}$ and later show that there is no
TIE transformation of $\alg_{V,S,T,\alpha}$ without a significant loss
in worst-case approximation for some value of $\alpha$.

Given $y \subset [n]$, we define 
\[ n_T(y) = |y \cap T| + |y \cap U| \]
and 
\[ n_S(y) = |y \cap S| + 2|y \cap V|. \]
That is, $n_T(y)$ is the number of elements of $y$ that lie in $T$, with elements of $U$ counted twice.  Likewise, $n_S(y)$ is the number of elements of $y$ that lie in $S$, with elements of $V$ counted thrice.

The algorithm $\alg_{V,S,T,\alpha}$ is then described as Algorithm \ref{alg.algvst.tie}.

\begin{algorithm}[h]
\label{alg.algvst.tie}
\KwIn{Subset $y \in [n]$ of agents with value $1$}
\KwOut{An allocation $\allocs \in \Feas_{V,S,T,\alpha}$}
\BlankLine
\uIf{ $n_S(y) \geq t$, $n_S(y) \geq \gamma|y|$, and $n_S(y) \geq n_T(y)$ }{
\Return{$\allocs_S^1$}
}
\uElseIf{ $n_T(y) \geq t$, $n_T(y) \geq \gamma|y|$, and $n_T(y) \geq n_S(y)$ }{
\Return{$\allocs_T^{\alpha}$}
}
\Else{
\Return{$\allocs_{[n]}^{\gamma}$}
}
\caption{Allocation Algorithm $\alg_{V,S,T,\alpha}$}
\end{algorithm} 

\subsection{Analysis}

In this section, we derive the key lemmas for the proof of Theorem~\ref{thm.tie.lowerbound}.
First, we bound the approximation factor of algorithm $\alg_{V,S,T,\alpha}$ for problem $\Feas_{V,S,T,\alpha}$.

\begin{lem}
$approx_{\Feas_{V,S,T,\alpha}}(\alg_{V,S,T,\alpha}) \geq \alpha/6$.
\end{lem}
\begin{proof}
Choose $y \subseteq [n]$ and consider the three cases for the output of $\alg_{V,S,T,\alpha}$.

\textbf{Case 1: $n_S(y) \geq t$, $n_S(y) \geq \gamma|y|$, and $n_S(y) \geq n_T(y)$.}  Our algorithm returns allocation $\allocs_{S}^1$ and obtains welfare at least $|y \cap S|$.  Note that 
\[|y \cap S| \geq \frac{1}{3}n_S(y) \geq \frac{1}{3}t\] 
and 
\[|y \cap S| \geq \frac{1}{3}n_S(y) \geq \frac{1}{3}n_T(y) \geq \frac{1}{3}|y \cap T|.\]

The allocation $\allocs_{T}^{\alpha}$ obtains welfare at most $\alpha(|y\cap T|+|T\backslash y|v) \leq \alpha(|y \cap T| + nv\gamma) \leq |y \cap T|+t \leq 6|y \cap S|$.  Note that here we used $|T| \leq n\gamma$, which follows since $r > \gamma^{-1}$.  

The allocation $\allocs_{[n]}^{\gamma}$ obtains welfare at most $\gamma|y| + t \leq 2n_S(y) \leq 6|y \cap S|$.  So we obtain at least a $1/6$-approximation in this case.

\textbf{Case 2: $n_T(y) \geq t$, $n_T(y) \geq \gamma|y|$, and $n_T(y) \geq n_S(y)$.}  Our algorithm returns allocation $\allocs_{T}^\alpha$ and obtains welfare at least $\alpha|y \cap T|$.  The same argument as case 1 shows that our approximation factor is at least $\alpha/6$ in this case.

\textbf{Case 3: $n_S(y) \leq t$ and $n_T(y) \leq t$.}  Our algorithm returns allocation $\allocs_{[n]}^\gamma$ for a welfare of at least $\gamma(|y| + v(n-|y|)) \geq t$.  The allocation $\allocs_S^1$ obtains welfare at most $|y\cap S| + t \leq n_S(y) + t \leq 2t$, and allocation $\allocs_T^\alpha$ obtains welfare at most $2\alpha t \leq 2t$.  So our approximation factor is at least $1/2$ in this case.

\textbf{Case 4: $n_S(y) \leq \gamma|y|$ and $n_T(y) \leq \gamma|y|$.}  Our algorithm returns allocation $\allocs_{[n]}^\gamma$ for a welfare of at least $\gamma(|y| + v(n-|y|)) \geq \gamma|y|$.  The allocation $\allocs_S^1$ obtains welfare at most $|y\cap S| + t \leq 2n_S(y) \leq 2\gamma|y|$, and allocation $\allocs_T^\alpha$ obtains welfare at most $|y \cap T| + t \leq 2\alpha\gamma|y| \leq 2\gamma|y|$.  So our approximation factor is at least $1/2$ in this case.
\end{proof}

Suppose now that $\alg'$ is any algorithm for problem
$\Feas_{V,S,T,\alpha}$ that is TIE. We will show that $\alg'$ is then
very restricted in the allocations it can return on inputs $y = V$ and
$y= U$. Furthermore, we note that if $\alg'$ has a good enough
approximation ratio, then its allocations on inputs $y = V$ and $y =
U$ are restricted further still. In particular, the optimal allocation
on both $V$ and $U$ is $\alloc_S^1$; so to obtain a good approximation
factor, on both $U$ and $V$, the algorithm should allocate a large
enough amount to agents in $U$. As any TIE transformation of $\alg$
is itself an algorithm for problem $\Feas_{V,S,T,\alpha}$, these
observations will later play a key role in our impossibility result.

\begin{claim}
\label{claim.nontruthful}
Suppose $\alg'$ is a truthful-in-expectation algorithm for problem $\Feas_{V,S,T,\alpha}$.   Then the expected allocation to each agent in $U$ must be at least as large in $\alg'(U)$ as in $\alg'(V)$.
\end{claim}
\begin{proof}
Take any set $W$ with $V \subseteq W \subseteq U$, $|W| = |V| +1$.
Then, on input $W$, the expected allocation to the agent in
$W\setminus V$ must not decrease.  Since all allocations are constant on $U$, this means that the expected allocation to each agent in $U$ must not decrease.  By the same argument, $\alg'$ returns an allocation at least this large for all $W$ such that $V \subseteq W \subseteq U$, and in particular for $W = U$.
\end{proof}

In light of these claims, our strategy for proving Theorem \ref{thm.tie.lowerbound} will be to show that a polytime transformation $\calt$ is unlikely to encounter the allocation $\allocs_T^{\alpha}$ during its sampling when the input is $V$, given that the sets $V$, $S$, and $T$ are chosen uniformly at random over all admissible tuples.  This means the transformation will be unable to learn the value of $\alpha$.  This is key since it prevents the transformation from using the value of $\alpha$ to appropriately randomize between the allocation of $\allocs_S^1$ and the essentially empty allocation of $\allocs_{[n]}^\gamma$ to acheive an effective allocation of $\alpha$ for agents in $U$ on input $V$ thereby satisfying the conditions of Claim~\ref{claim.nontruthful}.
Similarly, a transformation is unlikely to encounter the allocation $\allocs_S^1$ during its sampling on input $U$, and therefore can not satisfy Claim~\ref{claim.nontruthful} by allocating $1$ to agents in $U$ on input $U$.  


\begin{lem}
\label{lem.lem1}
Fix $V$ and $S$ satsifying the requirements of admissibility.  Then for any $y \subseteq [n]$, $\Pr_T[ \alg_{V,S,T,\alpha}(y) = x_T^\alpha ] \leq e^{-O(\frac{t}{r+1})}$, with probability taken over all choices of $T$ that are admissible given $V$ and $S$.
\end{lem}
\begin{proof}
Fix any $y$.  Write $n_V = |y\cap V|$, $n_S = |y\cap (S-V)|$, and $n_{*} = |y \cap ([n]-S)|$.  Note that $|y| = n_V + n_S + n_{*}$.  Define the random variables $m_U$ and $m_T$ by $m_U = |y \cap (U-V)|$ and $m_T = |y \cap (T\backslash S)|$.  

The event $[\alg_{V,S,T,\alpha}(y) = x_T^\alpha]$ occurs precisely if the following are true:
\begin{equation}
\label{eq.1a}
m_T + 2n_V + 2m_U \geq t,
\end{equation}
\begin{equation}
\label{eq.1b}
m_T + 2n_V + 2m_U \geq \gamma(n_V + n_S + n_{*}),
\end{equation}
\begin{equation}
\label{eq.1c}
m_T + 2n_V + 2m_U \geq n_S + 3n_V.
\end{equation}

We will show that the probability of these three inequalities being true is exponentially small.  To see this, note that \eqref{eq.1c} implies that $m_T + 2m_U \geq n_V$.  Thus, \eqref{eq.1a} implies that $m_T + 2m_U \geq t/3$, and hence $m_T + m_U \geq t/6$.  Now each element of $y$ counted in $n_S$ will count toward $m_U$ with probability $\frac{1}{r+1}$, and each element of $y$ counted in $n_{*}$ will count toward $m_T$ with probability $\frac{1}{r+1}$.  Since $t \gg r$, Chernoff bounds imply that with probability at least $1 - e^{-O(t/r)}$, we will have $n_{*} + n_S \geq \frac{r}{2}(m_T+m_U)$.  Then
\[ \frac{m_T + 2n_V + 2m_U}{n_V + n_S + n_{*}} < \frac{3(m_T + 2m_U)}{n_S + n_{*}} < \frac{12}{r} \ll \gamma \]
contradicting \eqref{eq.1b}.
\end{proof}

\begin{lem}
\label{lem.lem2}
Fix $U$ and $T$ satsifying the requirements of admissibility.  Then
for any $y \subseteq [n]$, $\Pr_S[ \alg_{V,S,T,\alpha}(y) = x_S^1 ]
\leq e^{-O(\frac{t}{r+1})}$, with probability taken over all choices
of $V$ and $S$ that are admissible given $U$ and $T$.
\end{lem}
\begin{proof}
Fix any $y$.  Write $n_U = |y\cap U|$, $n_T = |y\cap (T-U)|$, and $n_{*} = |y \cap ([n]-T)|$.  Note that $|y| = n_U + n_T + n_{*}$.  Define the random variables $m_V$ and $m_S$ by $m_V = |y \cap V|$ and $m_S = |y \cap (S\backslash T)|$.  

The event $[\alg_{V,S,T,\alpha}(y) = x_S^1]$ occurs precisely if the following are true:
\begin{equation}
\label{eq.2a}
m_S + n_U + 2m_V \geq t,
\end{equation}
\begin{equation}
\label{eq.2b}
m_S + n_U + 2m_V \geq \gamma(n_U + n_T + n_{*}),
\end{equation}
\begin{equation}
\label{eq.2c}
m_S + n_U + 2m_V \geq n_T + 2n_U.
\end{equation}

We will show that the probability of these three inequalities being true is exponentially small.  To see this, first note that we can assume $n_T = 0$, as this only loosens the requirements of the inequalities.  We then have that \eqref{eq.2c} implies $m_S + 2m_V \geq n_U$.  Thus, \eqref{eq.2a} implies that $m_S + 2m_V \geq t/2$, and hence $m_S + m_V \geq t/4$.  Now each element of $y$ counted in $n_U$ will count toward $m_V$ with probability $\frac{1}{r}$, and each element of $y$ counted in $n_{*}$ will count toward $m_S$ with probability $\frac{1}{r+1}$.  Since $t \gg r$, Chernoff bounds imply that with probability at least $1 - e^{O(t/r)}$, we will have $n_{*} + n_T \geq \frac{r}{2}(m_S+m_V)$.  Then
\[ \frac{m_S + n_U + 2m_V}{n_U + n_{*}} < \frac{4(m_S + m_V)}{n_U + n_{*}} < \frac{8}{r} \ll \gamma \]
contradicting \eqref{eq.2b}.
\end{proof}

\subsection{Proof of Main Theorem}

We can now set our parameters $t$, $r$, $\alpha$, and $\gamma$.  We will choose $t = n^{1/5}$, $r = n^{3/20}$, and $\gamma = n^{-2/20}$.  The values of $\alpha$ we will be considering are $1$ and $n^{-1/20}$.  Note that $t \gg r \gg \gamma^{-1} \gg \alpha^{-1}$ for each choice of $\alpha$.  Note also that $v = t\gamma^{-1} / n = n^{-14/20} \ll 1$.  

Our idea now for proving Theorem \ref{thm.tie.lowerbound} is that since the transformation can not determine the value of $\alpha$ on input $V$ (by Lemma \ref{lem.lem1}), and since it can not find the ``good'' allocation of $\allocs_S^1$ on input $U$ (by Lemma \ref{lem.lem2}), it must be pessimistic and allocate the minimum possible value of $\alpha$ to agents in $V$ on input $V$ in order to guarantee that the resulting allocation rule is TIE (by Claim \ref{claim.nontruthful}).  This implies a bad approximation on input $V$ and hence a bad worst-case approximation.  

\begin{proofof}{Theorem \ref{thm.tie.lowerbound}}
For each admissible $V, S,T$ and $\alpha \in \{1,n^{-1/20}\}$, write $\alg'_{V,S,T,\alpha}$ for
$\calt(\alg_{V,S,T,\alpha})$.  
Lemma \ref{lem.lem2} implies that, with all but exponentially small probability, $\alg'_{V,S,T,\alpha}$ will not encounter allocation $\allocs_S^1$ on input $U$. Thus, on input $U$, it can allocate at most $\alpha$ to each agent in $U$ in expectation (using the fact that $\alpha > \gamma$).  Then, since $\alg'_{V,S,T,\alpha}$ is incentive compatible, Claim \ref{claim.nontruthful} implies that $\alg'_{V,S,T,\alpha}$ must allocate at most $\alpha$ to each agent in $U$ on input $V$.

Now Lemma \ref{lem.lem1} implies that, with all but exponentially small probability, $\alg'_{V,S,T,\alpha}$ will not encounter allocation $\allocs_T^\alpha$ on input $V$, and thus is unaware of the value of $\alpha$ on input $V$.  Thus, to ensure incentive compatibility, $\alg'_{V,S,T,1}(V)$ must allocate at most $n^{-1/20}$ to each agent in $U$.  It therefore obtains a welfare of $|V|n^{-1/20} + t \leq n^{1/5} + n^{3/10} < n^{6/20}$, whereas a total of $|V| = n^{7/20}$ is possible with allocation $\allocs_S^1$.  Thus $\alg'_{V,S,T,1}$ has a worst-case approximation of $n^{-1/20}$, whereas $\alg_{V,S,T,1}$ has an approximation factor of $1/6$.  
\end{proofof}

We conclude with a remark about extending our impossibility result to TIE transformations under the weaker goal of preserving the expected social welfare under a given distribution $\dists$.  We would like to prove that, when agents' values are drawn according to a distribution $\dists$, any TIE transformation necessarily degrades the average welfare of some algorithm by a large factor.  The difficulty with extending our techniques to this setting is that a transformation may use the distribution $\dists$ to ``guess'' the relevant sets $V$ and $U$ (i.e.\ if the distribution is concentrated around the sets $V$ and $U$ in our construction).  One might hope to overcome this difficulty in our construction by hiding a ``true'' set $V$ (that generates a non-monotonicity) in a large sea of sets that could potentially take the role of $V$.  Then, if the transformation is unlikely to find a good allocation on input $U$, and unlikely to determine the value of $\alpha$ on any of these potential sets, and is further unable to determine which set is the ``true'' $V$, then it must be pessimistic and allocate the minimum potential value of $\alpha$ on any of these potential sets in order to guarantee truthfulness.  Unfortunately, our construction assumes that all allocations are constant on $U$, and this makes it difficult to hide a set $V$ while simultaneously making it difficult to discover a good allocation on input $U$.  We feel that, in order to make progress on this interesting open question, it is necessary to remove the assumption that all allocations are constant on $U$ which, in hand, seems to make it much more difficult to derive necessary conditions for a transformation to be TIE.

\section{An Impossibility Result for Makespan}
\label{sec.makespan}

We now consider an objective function, namely makespan, that differs from the social welfare objective in that it is not linear in agent values or allocations. Informally we show that black-box reductions for approximation algorithms for makespan are not possible even if we relax the notion of truthfulness to Bayesian incentive compatibility and relax the measure of performance to expected makespan, where both the notions are defined with respect to a certain {\em fixed} and {\em known} distribution over values. As in the previous section, our impossibility result hinges on the fact that the transformation is not aware of the feasibility constraint that an allocation needs to satisfy and can learn this constraint only by querying the algorithm at different inputs.

\subsection{Problem definition and main theorem}

We consider the following minimization problem in a Bayesian setting.  In this problem $n$ selfish machines (a.k.a. agents) are being allocated jobs.  Each agent has a private value $\vali$ representing its speed. If machine $i$ is allocated jobs with a total length $\alloci$, the load of machine $i$ is $\alloci / \vali$.  The makespan of allocation $\allocs$ to machines with speeds $\vals$ is the maximum load of any machine:
\[\msp(\allocs,\vals) = \max_i \frac\alloci\vali.\]
An instance of the (Bayesian) makespan problem is given by a feasibility constraint $\Feas$ and a distribution over values $\dists$; the goal is to map every value vector to allocations so as to minimize the expected makespan:
\[\expect[\vals\sim\dists]{\msp(\allocs(\vals),\vals)} \text{ subject to } \allocs(\vals)\in\Feas \text{ for all } \vals.\]
Given an algorithm $\alg$, we use $\Emsp(\alg)$ to denote its expected makespan.

Our main result is the following:

\begin{thm}
\label{thm.makespan.neg}
Let $n$ be large enough and $\calt$ be any black-box BIC transformation that makes at most $e^{n^{1/4} / 2}$ black-box queries to the given algorithm on each input.  There exists an instance $(\Feas,\dists)$ of the makespan problem and a deterministic algorithm $\alg$ such that $\calt(\alg)$ either returns an infeasible allocation with positive probability, or has makespan $\Emsp(\calt(\alg))= \Omega(n^{1/4})\Emsp(\alg)$.  Here $\dists$ is the uniform distribution over $\{1,\alpha\}^n$ for an appropriate $\alpha$ and is known to $\calt$.
\end{thm}

We note that the algorithm $\alg$ in the statement of Theorem \ref{thm.makespan.neg} is deterministic.  A BIC transformation $\calt$ must therefore degrade the makespan of some algorithms by a polynomially large factor even when we limit ourselves to deterministic algorithms. For simplicity of exposition, we prove a gap of $\Omega(n^{1/4})$, however, our construction can be tweaked to obtain a gap of $\Omega(n^{1/2-\delta})$ for any $\delta>0$.


\paragraph{Problem Instance.}

We now describe the problem instance $(\Feas,\dists)$ in more detail. Let $\alpha<n^{1/2}$ be a parameter to be determined later. As mentioned earlier, $\dists$ is the uniform distribution over $\{1,\alpha\}^n$. That is, every value (i.e. speed) $\vali$ is $1$ or $\alpha$ with equal probability.  There are $2n$ jobs in all, $n$ of length $\alpha$ and $n$ of length $1$.  Our feasibility constraint will have the property that each machine can be assigned at most one job.  So a valid allocation will set the allocation to each machine to a value in $\{0, 1, \alpha\}$.\footnote{A makespan assignment must allocate each job to a machine, but we will sometimes wish to specify an allocation in which not all jobs are allocated.  For ease of exposition, we will therefore assume that there is an extra agent with value $n(\alpha+1)$; this agent will always be allocated all jobs not allocated to any other agent.  Note that the load of this machine is always at most $1$.}

Of all such allocations (i.e. all $\allocs \in \{0, 1, \alpha\}^n$), all but one will be feasible.  This one infeasible allocation is thought of as a parameter of the problem instance.  Given $\allocs \in \{0,1,\alpha\}^n$, we will write $\Feas_\allocs$ as the set $\{0,1,\alpha\}^n\setminus\allocs$, and $\Gamma(\allocs)=(\Feas_\allocs,\dists)$ as the corresponding problem instance in which $\allocs$ is infeasible. We will use $\allocs_{bad}$ to denote the forbidden allocation in the remainder of this section.

\paragraph{The algorithm.}

We will first describe a randomized algorithm $\alg(\allocs_{bad})$ (Algorithm~\ref{alg.algvst.makespan} below) which we think of as an approximation algorithm for problem instance $\Gamma(\allocs_{bad})$.

\SetKw{KwGoto}{go to}
\SetKw{KwLine}{line}

\begin{algorithm}[h]
\label{alg.algvst.makespan}
\KwIn{Vector $\vals \in \{1,\alpha\}^n$}
\KwOut{An allocation $\allocs \neq \allocs_{bad}$}
\BlankLine
$H \leftarrow \{ i : \vali = \alpha \}$\;
\uIf{$\frac{1}{2}n - n^{3/4} \leq |H| \leq \frac{1}{2}n + n^{3/4}$}{
Choose set $S \subset H$ with $|S| = n^{-1/2}|H|$ uniformly at random\;
\lFor{$i \in S$}{$\alloci \leftarrow \alpha$}\;
\lFor{$i \in H\setminus S$}{$\alloci \leftarrow 0$}\;
\lFor{$i \not\in H$}{$\alloci \leftarrow 1$}\;
}
\Else{
Choose set $S \subset [n]$ with $|S| = n^{3/4}$ uniformly at random\;
\lFor{$i \in S$}{$\alloci \leftarrow \alpha$}\;
\lFor{$i \not\in S$}{$\alloci \leftarrow 0$}\;
}
\lIf{$\allocs = \allocs_{bad}$}{\KwGoto \KwLine 1}\;
\KwRet{$\allocs$}
\caption{Allocation Algorithm $\alg(\allocs_{bad})$}
\end{algorithm} 

We first note that $\alg(\allocs_{bad})$ must terminate.
\begin{claim}
For all $\vals$, algorithm $\alg(\allocs_{bad})$ terminates with probability $1$.
\end{claim}
\begin{proof}
This follows from noting that at least two distinct allocations can be chosen by $\alg(\allocs_{bad})$ on each branch of the condition on line $2$, so $\alg(\allocs_{bad})$ must eventually choose an allocation that is not $\allocs_{bad}$.
\end{proof}

We now use $\alg(\allocs_{bad})$ to define a set of deterministic algorithms\footnote{More precisely, we will define a set of deterministic allocation rules that map type profiles to allocations; in particular we will not be concerned with implementations of these allocation rules.}.  Let $D(\allocs_{bad})$ (or $D$ for short) denote the set of deterministic algorithms in the support of $\alg(\allocs_{bad})$.  That is, for every $\alg \in D$, $\alg(\vals)$ is an allocation returned by $\alg(\allocs_{bad})$ on input $\vals$ with positive probability for every $\vals \in \{1,\alpha\}^n$.  Moreover, for every combination of allocations that can be returned by $\alg(\allocs_{bad})$ on each input profile, there is a corresponding deterministic algorithm in $D$.

For any $\vals$, let $H(\vals) = \{ i : \vali = \alpha\}$ be the set of high speed agents.  Let $C$ denote the event that $\frac{1}{2}n - n^{3/4} \leq |H(\vals)| \leq \frac{1}{2}n + n^{3/4}$, over randomness in $\vals$.  We think of $C$ as the event that the number of high-speed agents is concentrated around its expectation.  We note the following immediate consequence of Chernoff bounds.

\begin{obs}
\label{obs.chernoff}
$\Pr_{\vals}[C] \geq 1 - 2e^{-n^{1/4}/4}.$
\end{obs}

This allows us to bound the expected makespan of each $\alg \in D$.

\begin{lem}
\label{lem.alg.makespan}
For each $\alg \in D$, $\Emsp(\alg)\le 1 + 2\alpha e^{-n^{1/4}/4}$ where the expectation is taken over $\vals$.
\end{lem}
\begin{proof}
If event $C$ occurs, then $\alg$ returns an allocation in which each agent $i$ with $\vali = 1$ receives allocation at most $1$.  Since each agent with $\vali = \alpha$ also receives allocation at most $\alpha$, we conclude that if event $C$ occurs then the makespan of $\alg(\vals)$ is at most $1$.  Otherwise, the makespan of $\alg(\vals)$ is trivially bounded by $\alpha$.  Since Observation \ref{obs.chernoff} implies that this latter case occurs with probability at most $2e^{-n^{1/4}/4}$, the result follows.
\end{proof}

\comment{

We next note that algorithm $\alg$ is not BIC, as long as $\alpha < n^{1/2}$.

\begin{lem}
Suppose $\alpha < n^{1/2}$.  Then, for each agent $i$, $\alloci(1) > \alloci(\alpha)$.
\end{lem}
\begin{proof}
Since $\vali = 1$ and $\vali=\alpha$ are equally likely, and the condition on line $2$ of the algorithm is symmetric with respect to the sizes of the high-speed and low-speed sets, we have that $\Pr[ C | \vali = 1 ] = \Pr[ C | \vali = \alpha ]$.  Call this probability $p$.

Consider the value of $\alloci(1)$.  If the condition on line $2$ evaluates true, which occurs with probability $p$, an agent with value $1$ receives allocation $1$.  On the other hand, if this condition evaluates false, then each agent with value $1$ receives allocation $\alpha$ with probability $n^{-1/4}$.  Thus  $\alloci(1) = p + (1-p)\alpha n^{-1/4}$.  

Next consider the value of $\alloci(\alpha)$.  If the condition on line $2$ evaluates true, then each agent with value $\alpha$ receives allocation $\alpha$ with probability $n^{-1/2}$.  If this condition evaluates false, then each agent again receives allocation $\alpha$ with probability $n^{-1/4}$.  Thus $\alloci(\alpha) = p\alpha n^{-1/2} + (1-p)\alpha n^{-1/4} < p + (1-p)\alpha n^{-1/4} = \alloci(1)$, since $\alpha < n^{1/2}$.
\end{proof}
}

\subsection{Transformation Analysis}

We now present a proof of Theorem~\ref{thm.makespan.neg}. Let $\calt$ denote a BIC transformation that can make at most $e^{n^{1/4} / 2}$ black-box queries to an algorithm for makespan.  Write $\calt(\alg)$ for the mechanism induced when $\calt$ is given black-box access to an algorithm $\alg$.

We first note that if $\calt(\alg)$ returns only feasible allocations with probability $1$, then it can only return an allocation that it observed during black-box queries to algorithm $\alg$.  This is true even if we consider only algorithms of the form $\alg \in D(\allocs_{bad})$ for some choice of $\allocs_{bad}$.

\begin{claim}
Suppose that for some $\alg \in D$, with positive probability, $\calt$ returns an allocation not returned by a black-box query to $\alg$.  Then there exists an algorithm $\alg'$ such that $\calt_{\alg'}$ returns an infeasible allocation with positive probability.
\end{claim}
\begin{proof}
Suppose that with positive probability $\calt_{\alg}$ returns the allocation $\allocs'$ on $\alg \in D(\allocs)$ without encountering it in a black box query to the algorithm $\alg$.  Then there exists $\alg' \in D(\allocs)$ such that $\alg'$ and $\alg$ agree on each input queried by $\calt$ in some instance where $\calt_{\alg}$ returns $\allocs'$, and furthermore $\alg'$ never returns allocation $\allocs'$ on any input.  Note, then, that $\calt_{\alg'}$ also returns allocation $\allocs'$ with positive probability.  But $\alg' \in D(\allocs')$,  so if we set $\allocs_{bad} = \allocs'$ then we conclude that $\calt_{\alg'}$ returns the infeasible allocation $\allocs'$ with positive probability.
%
\end{proof}

In the remainder of the analysis we assume that $\calt$ only returns observed allocations. 
We will now think of $\allocs_{bad}$ as being fixed, and $\alg$ as being drawn from $D(\allocs_{bad})$ uniformly at random.
Let $B$ denote the (bad) event, over randomness in $\vals$, $\calt$, and the choice of $\alg \in D$, that $\calt(\alg)$ returns an allocation with makespan $\alpha$.  Our goal will be to show that if $\calt(\alg)$ is BIC for every $\alg \in D$, then $\Pr[B]$ must be large; this will imply that the expected makespan of $\calt(\alg)$ will be large for some $\alg \in D$.

Intuitively, the reason that an algorithm $\alg \in D$ may not be truthful is because low-speed agents are often allocated $1$ while high-speed agents are often allocated $0$.  In order to fix such non-monotonicities, $\calt$ must either increase the probability with which $1$ is allocated to the high-speed agents, or increase the probability with which $0$ is allocated to the low-speed agents. To this end, let $U(\vals)$ be the event that, on input $\vals$, $\calt(\alg)$ returns an allocation $\allocs$ in which at least $n^{3/4}$ agents satisfy $\vali = 1$ and $\alloci = 0$.

As the following lemma shows, the event $U(\vals)$ is unlikely to occur unless $B$ also occurs. Then to fix the non-monotonicity while avoiding $B$, $\calt$ must rely on allocating $1$ more often to the high-speed agents.  However this would require $\calt$ to query $\alg$ on speed vectors $\vals'$ that are near-complements of $\vals$, and in turn imply a large-enough probability of allocating $\alpha$ to low-speed agents, i.e. event $B$. We now make this intuition precise.


\begin{lem}
For each $\vals$, $\Pr[U(\vals) \wedge \neg B | \vals] < (\ln 4)e^{-n^{1/4}/2}$.
\end{lem}
\begin{proof}
Fix input $\vals$, and suppose that event $U(\vals) \wedge \neg B$ occurs. Recall that $\calt$ only returns an allocation that $\alg$ outputs on a query $\vals'$. We will refer to a query of $\alg$ on input $\vals'$ as a \emph{successful} query, if it returns an $\allocs$ that satisfies $U(\vals) \wedge \neg B$. Let us bound the probability of a single query being successful. Let $t$ denote the number of agents with $\vali=1$. Then $U(\vals)$ implies $t\ge n^{3/4}$.

First, suppose that $\vals'$ does not satisfy the event $C$, that is, the number of high speed agents in $\vals'$ is far from its mean $n/2$. Then each of the $t$ agents has an 
$n^{-1/4}$ probability of being allocated $\alpha$ (taken over the choice of $\alg$ from $D$). The probability that none of the $t$ agents is allocated a load of $\alpha$ is at most $(1-n^{-1/4})^t < e^{-n^{1/2}}$.

On the other hand, suppose that $\vals'$ satisfies the event $C$. Then $U(\vals)$ implies that at least $t' \geq n^{3/4}$ agents satisfy $\vali = 1$ and $\vali' = \alpha$. 
In this case, each of these $t'$ agents has probability $n^{-1/2}$ of being allocated $\alpha$. The probability that none of them is allocated a load of $\alpha$ is at most $(1-n^{-1/2})^{t'} < e^{-n^{1/4}}$.

In either case, the probability that a single query is successful is at most $e^{-n^{1/4}}$.
\comment{
 This implies that $\calt$ returns an allocation $\allocs$ that satisfies the following properties:
\begin{enumerate}
\item $\allocs$ is returned by $\alg$ on some input $\vals'$ in which $t \geq n^{3/4}$ agents satisfy $\vali = 1$ and $\vali' = \alpha$. 
\item $\allocs$ satisfies $\msp(\allocs,\vals) < \alpha$.  
\end{enumerate}
We will refer to a query of $\alg$ on input $\vals'$ that returns such an $\allocs$ as a \emph{successful} query.  Let us first bound the probability of a single query being successful.  In a query $\vals'$ such that $t$ agents satisfy $\vali = 1$ and $\vali' = \alpha$, each of the $t$ agents has probability $n^{-1/2}$ of being allocated $\alpha$ if $\vals'$ satisfies event $C$, and probability $n^{-1/4}$ of being allocated $\alpha$ otherwise (with probabilities taken over the choice of $\alg$ from $D$).  Thus, in either case, each of the $t$ agents has probability at least $n^{-1/2}$ of being allocated $\alpha$, which in turn implies a makespan of $\alpha$. The probability that none of the $t$ agents is allocated a load of $\alpha$ is at most $(1-n^{-1/2})^t < e^{-n^{1/4}}$, which is therefore a bound on the probability that a query is successful.
}
Transformation $\calt$ can make at most $e^{n^{1/4}}/2$ queries on input $\vals$.  We will now bound the probability that any one of them is successful. First note that since $\alg$ is deterministic, we can assume that $\calt$ does not query $\alg$ more than once on the same input.  Furthermore, we can think of the choice of $\alg \in D$ as independently selecting the behaviour of $\alg$ for each input profile, so that the allocations returned by $\alg$ on different input profiles are independent with respect to the choice of $\alg$ from $D$.  We can therefore think of the $e^{n^{1/4}}/2$ queries as independent trials that are successful with probability at most $e^{-n^{1/4}}$.  Thus, the probability that at least one of these queries is successful is at most 
\begin{align*}
1 - (1-e^{-n^{1/4}})^{e^{n^{1/4}/2}} & = 1 - (1-e^{-n^{1/4}})^{e^{n^{1/4}}e^{-n^{1/4}/2}} \\
& < 1 - (1/4)^{e^{-n^{1/4}/2}} \\
& = 1 - (1/e)^{(\ln 4)e^{-n^{1/4}/2}} \\
& < (\ln 4)e^{-n^{1/4}/2}
\end{align*}
as required.
\end{proof}

We now consider the specific probabilities with which $\calt$ returns high or low allocations on high or low values. For agent $i$, value $\val$, and allocation $\alloc$, we will write $p_i^\alloc(\val) = \Pr_{\valsmi,\alg,\calt}[\alloci(\val,\valsmi) = \alloc]$.  That is, $p_i^\alloc(\val)$ is the probability that conditioned on agent $i$'s value being $\val$, $\calt(\alg)$ allocates $\alloc$ to the agent; Here the probability is over the values of the other agents, any randomness in $\calt$, and the choice of $\alg \in D$.

\begin{obs}
$\sum_i p_i^\alloc(\val) = 2\sum_i \Pr_{\vals}[(\vali = \val) \wedge (\alloci(\vals) = \alloc)]$.
\end{obs}

We can express the fact that $\calt$ satisfies BIC in terms of conditions on these probabilities (Lemma~\ref{lem.bic.prob.bound} below): either $p_i^0(1)$ should be large, i.e. low-speed agents get a low allocation, or one of $p_i^1(\alpha)$ and $p_i^\alpha(\alpha)$ should be large, i.e. high-speed agents get a high allocation. On the other hand, in Lemmas~\ref{lem.alpha.alpha}, \ref{lem.zero.one}, and \ref{lem.one.alpha} we show that on average over all agents, each of these probabilities is small if the probability of the bad event $B$ is small. The proofs of these lemmas are deferred to the end of this section. In Lemma~\ref{lem.bad.prob.bound} we put these results together to argue that $B$ occurs with a large probability.

\begin{lem}
\label{lem.bic.prob.bound}
Let $\allocs$ be the allocation rule of $\calt(\alg)$.  Then if $\alloci(1) \leq \alloci(\alpha)$, $p_i^0(1) < 1/3$, and $p_i^1(\alpha) < 1/3$, then $p_i^\alpha(\alpha) > 3 / \alpha$.
\end{lem}
\begin{proof}
Since $p_i^0(1) < \frac{1}{3}$, we have $\alloci(1) > \frac{2}{3}$.  Since $p_i^1(\alpha) < \frac{1}{3}$, we have $\alloci(\alpha) < \frac{1}{3} + p_i^\alpha(\alpha) \alpha$.  We conclude that $\frac{2}{3} < \frac{1}{3} + p_i^\alpha(\alpha) \alpha$ which implies the desired result.
\end{proof}


\begin{lem}
\label{lem.alpha.alpha}
$\frac{1}{2n}\sum_i p_i^\alpha(\alpha) \leq n^{-\frac{1}{2}} + \Pr[B] n^{-1/4} + 2e^{-n^{1/4}/4} n^{-1/4} + (\ln 4)e^{-n^{1/4}/2} n^{-1/4}.$
\end{lem}

\begin{lem}
\label{lem.zero.one}
$\frac{1}{2n}\sum_i p_i^0(1) \leq n^{-1/4} + \Pr[B] + (\ln 4)e^{-n^{1/4}/2}$.
\end{lem}

\begin{lem}
\label{lem.one.alpha}
$\frac{1}{2n}\sum_i p_i^1(\alpha) \leq 3n^{-1/4} + \Pr[B] + (\ln 4)e^{-n^{1/4}/2} + 2e^{-n^{1/4}/4}$.
\end{lem}


The above lemmas put together give a lower bound for $\Pr[B]$.  Set $\alpha = \frac{1}{4}n^{1/2}$.

\begin{lem}
\label{lem.bad.prob.bound}
$\Pr[B] \geq n^{-1/4} -  2e^{-n^{1/4}/4} - (\ln 4)e^{-n^{1/4}/2}$.
\end{lem}
\begin{proof}
We know that, for each $i$, either $p_i^0(1) \geq 1/3$, $p_i^1(\alpha) \geq 1/3$, or $p_i^\alpha(\alpha) > 3 / \alpha$.  So one of these inequalities must be true for at least one third of the agents, and hence one of the following must be true:
\[ \frac{1}{2n}\sum_i p_i^0(1) \geq 1/18 \]
\[ \frac{1}{2n}\sum_i p_i^1(\alpha) \geq 1/18 \]
\[ \frac{1}{2n}\sum_i p_i^\alpha(\alpha) \geq 1/2\alpha. \]
Suppose the first inequality is true.  Then by Lemma~\ref{lem.zero.one} we know 
\[ \Pr[B] \geq 1/18 - n^{-1/4} - (\ln 4)e^{-n^{1/4}/2} \]
which implies the desired result for sufficiently large $n$ (as the right hand side is at least a constant for large $n$, whereas $n^{-1/4} -  2e^{-n^{1/4}/4} - (\ln 4)e^{-n^{1/4}/2}$, from the statement of the lemma, vanishes as $n$ grows).

\noindent 
Suppose the second inequality is true.  Then we know from Lemma~\ref{lem.one.alpha}
\[ \Pr[B] \geq 1/18 - 3n^{-1/4} - (\ln 4)e^{-n^{1/4}/2} - 2e^{-n^{1/4}/4} \]
which again implies the desired result.

\noindent 
Finally, suppose the third inequality is true.  Then we know from Lemma~\ref{lem.alpha.alpha}
\[ n^{-\frac{1}{2}} + \Pr[B] n^{-1/4} + 2e^{-n^{1/4}/4} n^{-1/4} + (\ln 4)e^{-n^{1/4}/2} n^{-1/4} \geq 1/2\alpha \]
which implies (recalling $\alpha = \frac{1}{4}n^{1/2}$)
\[ \Pr[B] \geq (2 n^{-\frac{1}{2}} - n^{-\frac{1}{2}}) n^{1/4} - 2e^{-n^{1/4}/4} - (\ln 4)e^{-n^{1/4}/2}= n^{-1/4} - 2e^{-n^{-1/4}/4} - (\ln 4)e^{-n^{1/4}/2} \]
as required.
\end{proof}

We can now prove our main result.
\comment{
\begin{thm}
Suppose that $\calt$ is a BIC transformation for makespan.  Then for any $1/4 > 0$ and sufficiently large $n$, there exists a deterministic algorithm $\alg$ for which the expected makespan of $\alg$ is at most $1 + \frac{1}{2}n^{1/2} e^{-n^1/4/4}$ but the expected makespan of $\calt(\alg)$ is at least $\frac{1}{4}n^{1/4}$.
\end{thm}
}
\begin{proof}[Proof of Theorem~\ref{thm.makespan.neg}]
Write $\Pr[B\ |\ \alg]$ for the probability of event $B$ given that $\calt$ is given black-box access to algorithm $\alg$, with probability over the choice of input profile $\vals$.  Choose $\alg' \in \argmax_{\alg \in D}\{ \Pr[B\ |\ \alg] \}$.  Then in particular $\Pr[B\ |\ \alg] \geq \Pr[B]$, where recall that $\Pr[B]$ is the probability of event $B$ when $\alg$ is chosen uniformly at random from $D$.  

Recall that we set $\alpha=\frac 14 n^{1/2}$. By Lemma~\ref{lem.alg.makespan} the expected makespan of $\alg'$ is at most $1 + 2\alpha e^{-n^{1/4}/4} = 1 + \frac{1}{2} n^{1/2} e^{-n^{1/4}/4} <2$ for large enough $n$.  Using Lemma~\ref{lem.bad.prob.bound}, the expected makespan of $\calt_{\alg'}$ is at least 
\begin{align*}
1 + \alpha \Pr[B | \alg'] & \geq 1 + \alpha \Pr[B] \\
& \geq 1 + \alpha(n^{-1/4} -  2e^{-n^{1/4}/4} - (\ln 4)e^{-n^{1/4}/2}) \\
& = 1 + \frac{1}{4} n^{1/4} - \frac{1}{2}n^{1/2} e^{-n^{1/4}/4} - \left(\frac{\ln 4}{4}\right)n^{1/2} e^{-n^{1/4}/2} \\
& \geq \frac{1}{4}n^{1/4}
\end{align*}
as required.
\end{proof}

\paragraph{Proofs of bounds on the allocation probabilities.} To conclude the analysis, we now present proofs of Lemmas~\ref{lem.alpha.alpha}, \ref{lem.zero.one}, and \ref{lem.one.alpha}.

\begin{proof}[Proof of Lemma~\ref{lem.alpha.alpha}]
Let us first condition on the event $\neg B \wedge \neg U(\vals) \wedge C$. That is, we consider the output of $\calt(\alg)$ on a value vector $\vals$ that satisfies the concentration event $C$, and further assume that the makespan of $\calt(\alg)$ is small ($\neg B$) and few agents with $\vali=1$ have a $0$ allocation ($\neg U(\vals)$). 

$C$ implies that many of the agents (at least $\frac 12 n - n^{3/4}$) in $\vals$ are low-speed agents. Along with $\neg B$ and $\neg U(\vals)$ this implies that most of these agents have an allocation of $1$; Call this set of agents $L$. Then $|L|>\frac 12 n - 2n^{3/4}$. In particular $L$ is non-empty. Now suppose that $\calt(\alg)$ returns an allocation $\allocs$ that is returned by $\alg$ on input $\vals'$. Then, since $L$ is non-empty, $\vals'$ must satisfy the condition on line $2$ of $\alg(\allocs_{bad})$ (as this is the only way in which any agent can be allocated $1$, regardless of the choice of $\alg \in D$).  This implies that at most $n^{1/2}$ agents get an allocation of $\alpha$ because $S$ is of size at most $n^{1/2}$.

We conclude that for fixed $\vals$ satisfying $C$ and conditioning on $\neg B \wedge \neg U(\vals)$,
\[ \frac{1}{n}\sum_i \Pr[(\vali = \alpha) \wedge (\alloci(\alpha,\valsmi) = \alpha)] \leq \frac{1}{n}n^{\frac{1}{2}} = n^{-\frac{1}{2}}.\]



For the case that events $B$, $\neg C$, or $U(\vals)$ occur, we note that every allocation returned by $\calt(\alg)$ allocates $\alpha$ to at most $n^{3/4}$ agents.  So, conditioning on either of these events, we have 
\[ \frac{1}{n}\sum_i \Pr[(\vali = \alpha) \wedge (\alloci(\alpha,\valsmi) = \alpha)] \leq n^{-1/4}. \]
We conclude, taking probabilities over all $\vals$, that
\begin{align*}
\frac{1}{2n}\sum_i p_i^0(1) & \leq n^{-\frac{1}{2}} + \Pr[B] n^{-1/4} + \Pr[\neg C] n^{-1/4}  + \Pr[U(\vals)\wedge \neg B] n^{-1/4}\\ 
& \leq n^{-\frac{1}{2}} + \Pr[B] n^{-1/4} + 2e^{-n^{1/4}/4} n^{-1/4} + (\ln 4)e^{-n^{1/4}/2} n^{-1/4}
\end{align*}
as required.
\end{proof}

\begin{proof}[Proof of Lemma~\ref{lem.zero.one}]
For each input vector $\vals$, either event $\neg U(\vals)$ occurs, event $B$ occurs, or event $U(\vals) \wedge \neg B$ occurs. Event $\neg U(\vals)$ by definition gives us a bound on the number of agents with value $1$ that receive an allocation of $0$. 
So conditioning on this event (and keeping $\vals$ fixed) we have 
\[ \frac{1}{n}\sum_i \Pr[(\vali = 1) \wedge (\alloci(1,\valsmi) = 0)] \leq \frac{1}{n}n^{3/4} = n^{-1/4}. \]  
Thus, taking probabilities over all $\vals$, we have
\[\frac{1}{2n}\sum_i p_i^0(1) \leq n^{-1/4} + \Pr[B] + \Pr_{\vals}[U(\vals) \wedge \neg B | \vals] \leq n^{-1/4} + \Pr[B] + (\ln 4)e^{-n^{1/4}/2}\]
as required.
\end{proof}

\begin{proof}[Proof of Lemma~\ref{lem.one.alpha}]
Let us first fix $\vals$ and condition on the event $\neg B \wedge \neg U(\vals) \wedge C$.  Suppose that $\calt(\alg)$ returns an allocation $\allocs$ that is returned by $\alg$ on input $\vals'$.  

As in the proof of Lemma \ref{lem.alpha.alpha}, event $\neg B \wedge \neg U(\vals) \wedge C$ implies that $\calt(\alg)$ returns an allocation in which some agents are allocated $1$.  We therefore conclude that $\vals'$ satisfies the condition on line 2 of $\alg(\allocs_{bad})$.  This implies 
\[-2n^{3/4} \leq |H(\vals)| - |H(\vals')| \leq 2n^{3/4}.\]
Furthermore, $\neg U(\vals) \wedge \neg B$ implies that $|H(\vals') \backslash H(\vals)| \leq n^{3/4}$. Combining with the inequalities above, we conclude that $|H(\vals) \backslash H(\vals')| \leq 2n^{3/4} + n^{3/4}$.  Note that this is a bound on the number of agents such that $\vali = \alpha$ and $\vali' = 1$, which is also a bound on the number of agents such that $\vali = \alpha$ and $\alloci = 1$.

We conclude that, conditioning on event $\neg B \wedge \neg U(\vals) \wedge C$ and keeping $\vals$ fixed, we have
\[ \frac{1}{n}\sum_i \Pr[(\vali = \alpha) \wedge (\alloci(\alpha,\valsmi) = 1)] \leq \frac{1}{n}(3n^{3/4}). \]  
Thus, taking probabilities over all $\vals$ and all events, we have
\begin{align*}
\frac{1}{2n}\sum_i p_i^0(1) & \leq 3n^{-1/4} + \Pr[B] + \Pr_{\vals}[U(\vals) \wedge \neg B | \vals] + \Pr[\neg C] \\
& \leq 3n^{-1/4} + \Pr[B] + (\ln 4)e^{-n^{1/4}/2} + 2e^{-n^{1/4}/4}
\end{align*}
as required.
\end{proof}

\section{Additive objective functions}
\label{sec.additive}

In Section \ref{sec.makespan} we showed that no BIC approximation-preserving transformations are possible for the makespan objective. One of the properties of the social welfare objective that allows a BIC transformation where one cannot exist for makespan is that the objective function is additive across agents. This allows a transformation to focus on each agent individually while taking an aggregate view over other agents and preserving the performance with respect to the respective component of the objective function alone. \cite{HL-10} and \cite{HKM-11} formalize this idea as follows: for each agent $i$ they construct a mapping $g_i$ from the value space of $i$ to itself, and on input $\vals$ return the output of the algorithm on $g(\vals)=(g_1(\vali[1]), g_2(\vali[2]), \cdots)$. The mappings $g_i$ ensure the following three properties: 
\begin{enumerate}
\item[(P.1)] the mapping preserves the distribution over values of $i$, 
\item[(P.2)] the expected allocation of agent $i$ upon applying the mapping, i.e. $\alloci(g_i(\vali))$, is monotone non-decreasing in $\vali$, and,
\item[(P.3)] the contribution of agent $i$ to the overall social welfare is no worse than in the original algorithm. 
\end{enumerate}
The benefit of this approach is that if each agent's value space is single-dimensional or well structured in some other way, the computational problem of finding such a mapping becomes easy.

Given this construction, it is natural to ask whether there are other objectives that are additive across agents and for which such a per-agent ironing procedure works. We show in this section that for almost any objective other than social welfare, such an approach cannot work. In particular, given an objective satisfying some mild properties, we construct an instance such that for any mapping $g_i$ from agent $i$'s value space to itself that satisfies properties (P.1) and (P.2) above, property (P.3) fails to hold by an arbitrarily large factor.

We focus first on maximization problems. Note that for an objective function of the form $\max \E_{v}[\sum_i \alloci(\vals)h_i(\vali)]$ where $h_i$s are non-decreasing functions, the approach of \cite{HL-10} and \cite{HKM-11} works as-is to give a BIC approximation preserving transformation. In the sequel, we consider objectives of the form $\max \E_{v}[\sum_i h_i(\alloci(\vals))\vali]$ where $h_i$ is an arbitrary non-linear continuous function.

\begin{thm}
  Consider the objective $\max \E_{v}[\sum_i h_i(\alloci(\vals))\vali]$ where each $h_i$ is an arbitrary increasing function. Suppose that there exists an agent $i$ for which $h_i$ is a continuous super-linear function (i.e. $h_i(x)=\omega(x)$) or a continuous sub-linear function (i.e. $h_i(x)=o(x)$). Then for any $\epsilon\in (0,1)$, there exists a distribution over agent values and an algorithm $\alg$ such that any transformation that performs a per-agent ironing of the allocation function of $\alg$ achieving properties (P.1) and (P.2) above, must violate property (P.3) by a factor of $\Omega(1/\epsilon)$.
\end{thm}
\begin{proof}
  We focus on a single agent $i$ and drop the subscript $i$ to improve readability. Our algorithm makes non-zero allocations only to agent $i$ so that the contribution of other agents to the objective is $0$. Let $h$ be the corresponding continuous increasing function and assume wlog that $h(0)=0$ and $h(1)=1$. In the remainder of the proof we assume that $h$ is super-linear. The proof for the sub-linear case is similar. 

With respect to this agent, our goal is to maximize the objective $E[h(\alloc(\val))\val]$. Fix any $\epsilon>0$ and consider the following instance and algorithm. The agent's value is $0$ with probability $1-\epsilon$ and $1$ with probability $\epsilon$. At $0$, the algorithm always allocates an amount $1+\epsilon'$. At $1$, the allocation is $0$ with probability $1-1/k$ and $k$ with probability $1/k$. Here we pick $\epsilon'>0$ and $k$ such that $h(1+\epsilon')< 1/(1-\epsilon)$ and $h(k)> k/\epsilon$. The existence of $\epsilon'$ and $k$ follows from the fact that $h$ is continuous and superlinear.

Now, the expected allocation at $1$ is $1$, so in order to produce a BIC output, the mapping $g$ must map each of the values to the other with some probability. Suppose that $0$ gets mapped to $1$ with probability $z/(1-\epsilon)$ for some $z$. Then, to preserve the distribution over values, we must have $z\le\epsilon$ and $1$ must get mapped to $0$ with probability $z/\epsilon$. 

How large does $z$ have to be to fix the non-monotonicity? The new expected allocation at $0$ is $z/(1-\epsilon) + (1-z/(1-\epsilon))(1+\epsilon')$, while the new expected allocation at $1$ is $(1-z/\epsilon)+z/\epsilon(1+\epsilon')$. Setting the former to be no larger than the latter, and rearranging terms, we get 
\[z/(1-\epsilon)+z/\epsilon\ge 1\]
implying $z\ge \epsilon(1-\epsilon)$.

Let us now compute the objective function value. The original objective function value is $\epsilon h(k)/k$. The new objective function value is given by
\begin{align*}
\epsilon ( z/\epsilon h(1+\epsilon') + (1-z/\epsilon) h(k)/k ) & < 
z/(1-\epsilon) + (\epsilon-z) h(k)/k \\
& < h(k)/k (\epsilon-z+\epsilon z/(1-\epsilon))\\
& = h(k)/k (\epsilon-(1-2\epsilon)z/(1-\epsilon))\\ 
& < 2\epsilon^2 h(k)/k
\end{align*}
Here the first inequality follows from $h(1+\epsilon')< 1/(1-\epsilon)$, the second from $1<\epsilon h(k)/k$ and the fourth from $z\ge \epsilon(1-\epsilon)$. This implies that any mapping $g$ that satisfies properties (P.1) and (P.2) must violate property (P.3) by a factor of at least $1/2\epsilon$. 

A similar example can be constructed for sub-linear continuous $h$, and we skip the details.
\end{proof}

Next we consider minimization problems of the form $\min \E_{v}[\sum_i h_i(\alloci(\vals))h'_i(\vali)]$ where $h_i$s are non-decreasing functions and $h'_i$s are non-increasing functions. Once again, if there exists an $i$ such that $h_i$ is non-linear, we get a gap.

\begin{thm}
  Consider the objective $\max \E_{v}[\sum_i h_i(\alloci(\vals))h'_i(\vali)]$ where each $h_i$ is an arbitrary increasing function and each $h'_i$ is an arbitrary continuous decreasing function. Suppose that there exists an agent $i$ for which $h_i$ is a continuous super-linear function (i.e. $h_i(x)=\omega(x)$) or a continuous sub-linear function (i.e. $h_i(x)=o(x)$). Then for any $\eps>0$, there exists a distribution over agent values and an algorithm $\alg$ such that any transformation that performs a per-agent ironing of the allocation function of $\alg$ achieving properties (P.1) and (P.2) above, must violate property (P.3) by a factor of $\Omega(1/\eps)$.
\end{thm}
\begin{proof}
  Once again we focus on the agent $i$ and present the proof for the case where the function $h$ (the subscript $i$ being implicit) is continuous and super-linear. The proof for the sub-linear case is similar. Assume without loss of generality that $h(0)=0$ and $h(1)=1$. Let $k\ge 1$ be such that $h(k)>k/\eps$. Let $v_1=h'^{-1}(1)$ and $v_2=h'^{-1}(\eps/k)$. Agent $i$'s value distribution is uniform over $\{v_1,v_2\}$. 

The algorithm $\alg$ is defined as follows. At $v_1$, the algorithm returns $\alloc_1$ with $h(\alloc_1)=1+\epsilon$. Note that $\alloc_1>1$. At $v_2$, the algorithm returns $k$ with probability $1/k$ and $0$ otherwise. The expected allocation is $1$, and therefore the allocation is non-monotone. Suppose that the tranformation maps $v_1$ to $v_2$ with probability $z$ and vice versa. Then it is easy to see that $z> 1/2$.

Now let us compute the objective function value. The objective function value of the original algorithm is $1/2(1+\eps) + 1/2 \eps/k (h(k)/k) = 1/2 + \eps/2(1+h(k)/k^2)< 1+\eps h(k)/2k^2$. We can bound from below the objective function value of the transformed mechanism by considering the term corresponding to value $v_1$ when the allocation is $k$ (an event that happens with probability $1/2$ times $z$ times $1/k$). The new objective function value is therefore at least
\begin{align*}
z h(k)/2k & \ge h(k)/4k \\
& \ge 1/8\eps + h(k)/16k^2 = 1/8\eps(1+\eps h(k)/2k^2)
\end{align*}
Here the second inequality follows by using $h(k)>k/\eps$ and $k>1/2$.  This implies that any transformation that satisfies properties (1) and (2) must violate property (3) by a factor of at least $1/8\epsilon$. 
\end{proof}

To conclude this section we note that for minimization problems with objectives of the form $\E_{v}[\sum_i \alloci(\vals) h'_i(\vali)]$ where $h'_i$s are decreasing functions, an approach similar to the ironing approach of \cite{HKM-11} gives a BIC approximation-preserving transformation. In particular, if the type space for each agent is finite then we can find a mapping by finding the min-cost perfect matching where edge costs between $\vali$ and $\vali'$ are $h'_i(\vali)\E_{\valsmi}[\alloci(\vali',\valsmi)]$.

\bibliographystyle{plain}
\bibliography{auctions}


\end{document}